\documentclass[conference]{IEEEtran}
\IEEEoverridecommandlockouts
\usepackage{cite}
\usepackage{amsmath,amssymb,amsfonts}
\usepackage{algorithmic}
\usepackage{textcomp}
\usepackage{xcolor}
\usepackage{booktabs}
\usepackage[pdftex]{graphicx}
\usepackage{url}
\DeclareGraphicsExtensions{.pdf,.png,.jpg}

\def\BibTeX{{\rm B\kern-.05em{\sc i\kern-.025em b}\kern-.08em
    T\kern-.1667em\lower.7ex\hbox{E}\kern-.125emX}}
\begin{document}

\title{AWE: Adaptive Agents for Dynamic Web Penetration Testing
}
\IEEEoverridecommandlockouts

\author{\IEEEauthorblockN{Akshat Singh Jaswal\IEEEauthorrefmark{1}}
\IEEEauthorblockA{Stux Labs\\
akshat@stuxlabs.com}
\thanks{\IEEEauthorrefmark{1}The authors contributed equally to this work.}
\and
\IEEEauthorblockN{Ashish Baghel\IEEEauthorrefmark{1}}
\IEEEauthorblockA{Stux Labs\\
ashish@stuxlabs.com}}

\makeatletter\def\@IEEEpubidpullup{6.5\baselineskip}\makeatother
\IEEEpubid{\parbox{\columnwidth}{
		{\fontsize{7.5}{7.5}\selectfont Workshop on LLM Assisted Security and Trust Exploration (LAST-X) 2026 \\
			27 February 2026, San Diego, CA, USA \\
			ISBN 978-1-970672-05-3 \\
			https://dx.doi.org/10.14722/last-x.2026.23037  \\
			www.ndss-symposium.org}
}
\hspace{\columnsep}\makebox[\columnwidth]{}}

\maketitle
\begin{abstract}
Modern web applications are increasingly produced through AI-assisted development and rapid no-code deployment pipelines, widening the gap between accelerating software velocity and the limited adaptability of existing security tooling. Pattern-driven scanners fail to reason about novel contexts, while emerging LLM-based penetration testers rely on unconstrained exploration, yielding high cost, unstable behavior, and poor reproducibility.

We introduce AWE, a memory-augmented multi-agent framework for autonomous web penetration testing that embeds structured, vulnerability-specific analysis pipelines within a lightweight LLM orchestration layer. Unlike general-purpose agents, AWE couples context aware payload mutations and generations with persistent memory and browser-backed verification to produce deterministic, exploitation-driven results.

Evaluated on the 104-challenge XBOW benchmark, AWE achieves substantial gains on injection-class vulnerabilities - 87\% XSS success (+30.5\% over MAPTA) and 66.7\% blind SQL injection success (+33.3\%) - while being much faster, cheaper, and more token-efficient than MAPTA, despite using a mid-tier model (Claude Sonnet 4) versus MAPTA’s GPT-5. MAPTA retains higher overall coverage due to broader exploratory capabilities, underscoring the complementary strengths of specialized and general-purpose architectures.
Our results demonstrate that architecture matters as much as model reasoning capabilities: integrating LLMs into principled, vulnerability-aware pipelines yields substantial gains in accuracy, efficiency, and determinism for injection-class exploits. The source code for AWE is available at: \url{https://github.com/stuxlabs/AWE}\end{abstract}


\section{Introduction}
The increasing popularity of AI assisted software development and the limited adaptablity of traditional security tools have created a widening gap in the web security landscape. Most notably recent developement trends of no-code platforms, automated code-generation assistants, and rapid deployment pipelines allow web applications to be made by developers with limited security expertise. This broadens the attack surface significantly while existing security tooling remain stuck in pattern based detections and lack genuine reasoning capabilities.

Recent OWASP Top 10 data shows that every major category of web weakness ranging from injection flaws to access control failures and server-side request manipulation continues to appear across most real world applications\cite{owasptop10}. Despite ongoing advancements in secure development practices, these vulnerability classes remain persistent. The widening gap between accelerated development and static defensive capabilities has created a massive challenge for modern web security assessment.

To address this growing mismatch, we introduce AWE (Adaptive Web Exploitation Framework), a memory-augmented multi-agent penetration testing system designed for autonomous, intelligent, and transparent vulnerability discovery. AWE aims to bridge the gap between traditional scanners and general-purpose LLM agents by combining domain-specific exploitation logic with large language models, enabling targeted, explainable, and scalable vulnerability discovery in modern web applications.

\section{Threat Model}
\subsection{System Model}
We consider an automated black box vulnerability discovery system that interacts with modern web applications through standard HTTP interfaces. The system exercises application endpoints using both GET and POST requests and uses parameter placement to explore multiple input channels. 
The target applications resemble modern web applications that expose parameterized HTTP endpoints and perform server-side processing using common frameworks (PHP, Python, Node.js, Java). These applications may incorporate input validation, output encoding, and application-layer firewalls.
The system has no privileged visibility into source code, runtime logs, or internal application state. All observations arise solely from HTTP responses and timing behavior. The attacker’s automation maintains a persistent memory across probes, enabling adaptive exploration.
\subsection{Attacker Capabilities and Goals}
We assume an automated adversary that interacts with the application strictly through black-box HTTP requests. The attacker is realistic, constrained, and possesses the following capabilities:
\begin{itemize}
    \item \textbf{Black-box interaction.}
    The attacker can craft arbitrary HTTP requests and observe responses, but lacks
    source-code access, server configuration details, or a privileged capabilities.

    \item \textbf{Authenticated probing.}
    When available through benign registration or low-privilege accounts, the attacker may
    authenticate to explore additional endpoints and restricted input
    channels.

    \item \textbf{LLM-assisted input generation.}
    The attacker employs commercial LLM APIs to synthesize context-aware payloads and
    adapt strategies based on prior observations, subject to cost-bounded operation. The LLM serves as a flexible generator of candidate attack
    inputs.

    \item \textbf{Time-bounded evaluation.}
    Each target endpoint is probed under a strict temporal budget ($\leq 10$ minutes),
    reflecting practical constraints imposed by rate limiting, detection risk, and LLM API
    costs.
\end{itemize}

The attacker’s goals are to identify injection-class vulnerabilities throughout controlled manipulation of inputs, and exploit abnormalities in application behavior (e.g., response timing, error structure, output differences) to infer server-side faults.

\subsection{Trust Relationship}
We assume the target application stack is uncompromised and behaves according to its implementation, although it may contain vulnerabilities. The hosting infrastructure and network fabric are also considered trustworthy from a security standpoint, providing no privileged access to the adversary.
All attacker-controlled inputs (parameters, headers, cookies, request bodies) are considered potentially malicious, and dynamic content originating from the client side may also serve as a vehicle for exploit construction
\begin{figure*}[!t]
\centering
\includegraphics[width=\columnwidth]{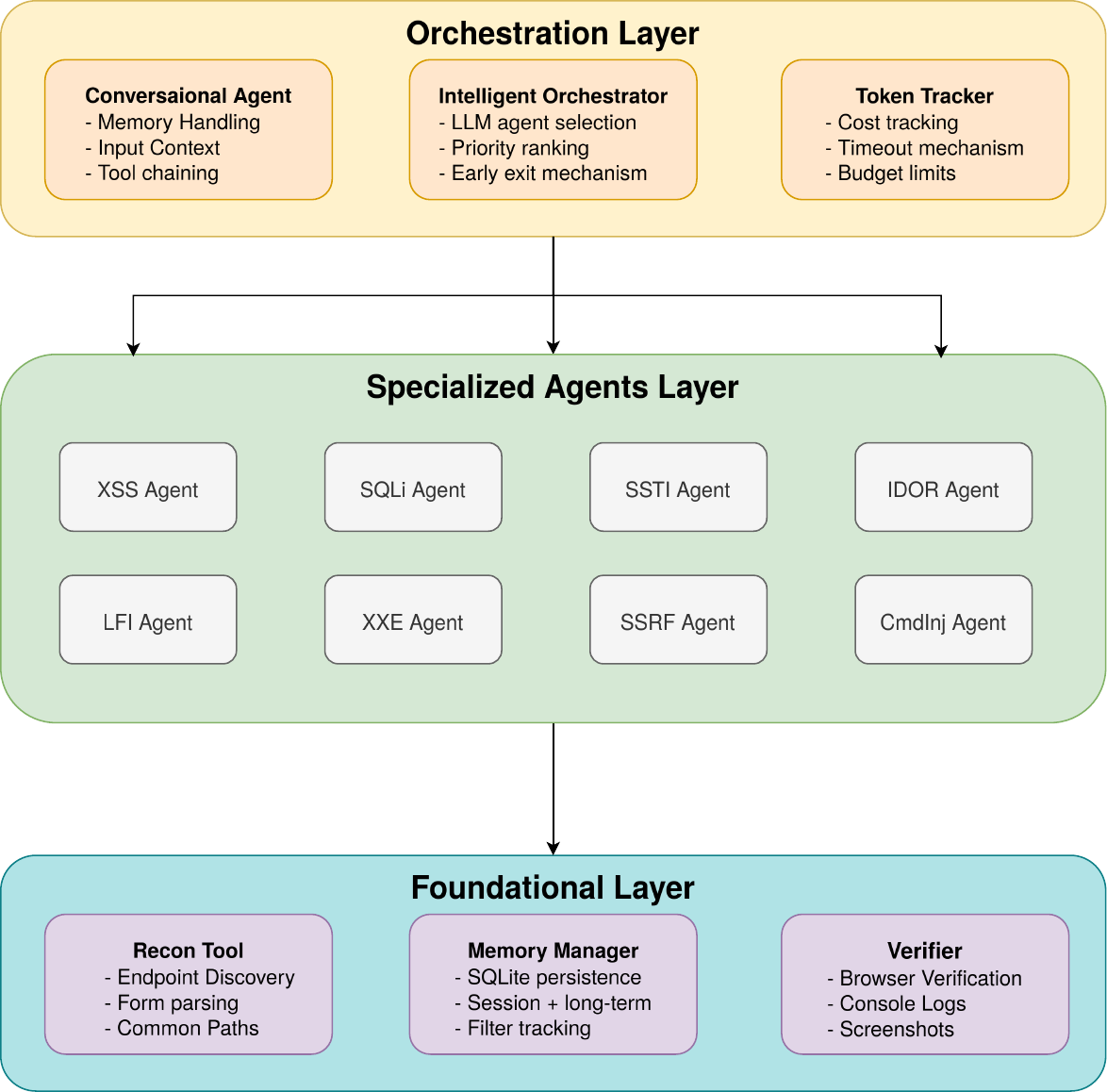}
      \caption{AWE system architecture. }
\label{fig:architecture}
\end{figure*}

\subsection{Scope}
Our work focuses on injection-centric vulnerabilities that can be discovered solely through black-box manipulation of application inputs. Within this scope, we consider attacks that exploit improper handling of attacker-controlled data across a broad spectrum of server-side interfaces. These include vulnerabilities such as cross-site scripting, SQL injection in its various forms, server-side template injection across widely used templating engines, command injection, file inclusion and path traversal, XML external entity expansion, server-side request forgery, and unauthorized object access when valid credentials are available. The scope therefore encompasses vulnerabilities whose exploitability emerges from observable differences in application behavior under controlled input perturbation.

Out of scope are vulnerabilities that cannot be meaningfully exercised or detected through black-box interaction alone. We do not consider network-level or protocol-level attacks, cryptographic weaknesses, or business-logic flaws that require semantic domain knowledge or multi-step reasoning beyond observable request–response behavior. Our focus is strictly on vulnerabilities that arise from input processing behavior accessible to a realistic, resource-constrained adversary operating through standard HTTP interfaces.

\section{Background and Related Work}

\subsection{Traditional Automated Vulnerability Scanning}
Dynamic Application Security Testing tools remain the popular automated method for identifying security flaws in modern web applications. Commercial systems such as 
Burp Suite~\cite{burp}, as well as open-source tools like OWASP ZAP~\cite{zap}, 
Nuclei~\cite{nuclei}, and sqlmap~\cite{sqlmap}, rely primarily on signature-driven payload databases combined with heuristic pattern matching. These tools excel at detecting well-understood classes of injection vulnerabilities by replaying curated payloads across various input vectors but this strict pattern matching also embeds inherent limitations.

One significant flow is that  signature and template based scanners are static and they cannot synthesize novel payloads or mutate attack strategies when confronted with nonstandard sanitization, application specific input handling, or adaptive WAFs. Also, the rigidity of pattern matching leads to false positives when benign behaviors resemble known signatures, and false negatives when exploitation requires multi-step probing or contextual reasoning. \cite{b1}. Specialized tools such as sqlmap for SQL injection show excellent domain-specific performance but lack generality across heterogeneous vulnerability families and vulnerablities with dependent chaining. Collectively, these limitations highlight the difficulty of expressing dynamic attack reasoning within static scanners.

\subsection{LLM-Based Penetration Testing Systems}
Large language models have recently motivated systems that apply natural language to security assessment. PentestGPT~\cite{pentestgpt} was the first well crafted attempt that proved LLMs can support human testers by structuring workflows, suggesting reconnaissance strategies, and making exploit logic. Although impactful, these systems function primarily as assistive agents: humans maintain the memory, perform validation, and execute tools.
Subsequent research has explored autonomous operation through multi-agent orchestration. 
Frameworks such as AutoPT~\cite{autopt}, AutoAttacker~\cite{autoattacker}, CAI~\cite{cai}, 
and related multi-agent LLM systems~\cite{mimicking, mapta} couple LLM-driven controllers 
with command execution environments and reconnaissance tooling. These approaches automate 
selected penetration testing phases, but typically rely on unspecialized reasoning models 
and lack persistent memory for tracking authentication status, filter behavior, or previously 
attempted payloads, features essential for complex injections.
MAPTA~\cite{mapta} represents a significant advancement in autonomous LLM driven penetration testing. It employs a three role multi-agent architecture in which a Coordinator agent performs high-level planning, Sandbox agents execute commands and scripts within an isolated per job Docker environment, and a Validation agent converts candidate exploits into verified proof-of-concepts through concrete execution. By coupling LLM-based reasoning with structured tool orchestration and evidence-gated PoC validation, MAPTA demonstrates that fully autonomous end-to-end web exploitation is feasible and establishes a strong baseline for agent-driven security testing.

\subsection{Architectural Gaps in Existing Systems}
Despite their individual strengths, both traditional scanners and existing LLM-based penetration testing systems share several fundamental limitations.
Although LLM-based agents technically receive server-side feedback, they lack the domain-specific exploitation reasoning required to interpret that feedback and transform it into effective payload evolution. Practical exploitation often depends on subtle details (filter ordering, encoding quirks, type coercion behavior, template engine semantics, multi-parameter interactions etc.) that general-purpose LLM reasoning does not reliably model. As a result, existing systems tend to generate a small set of generic payloads, fail to recognize why they were blocked, and prematurely abandon the search rather than performing the iterative probing needed to infer sanitization logic or craft targeted bypasses. 

Furthermore, most architectures do not maintain the rich contextual state necessary for multi-step exploitation, such as tracking which payload variants were attempted, how filter behavior has changed across requests, or which response features signal partial progress. Without structured memory and explicit state modeling, agents cannot build the multi-hop reasoning chains required for difficult injection classes. 

Finally, the absence of domain-specialized probing techniques such as type confusion probing, template context shifting, timing-based inference, or controlled syntax fragmentation limits the ability of existing systems to move beyond superficial exploitation attempts.

These gaps reflect a broader challenge: current tools combine feedback and autonomous reasoning but lack the specialized, stateful, and iterative mechanisms necessary to convert raw feedback signals into precise, context-aware exploit generation.

\section{System Design}
AWE is designed as an autonomous web exploitation system that integrates reconnaissance, domain specialized vulnerability analysis, and adaptive LLM reasoning under explicit resource constraints. Its architecture uses global orchestration to orchestrate  vulnerability specific logic and has shared memory to enable systematic exploration of an application’s attack surface simultaneously ensuring that each component operates with clear and specific responsibilities. 
AWE consists of three architectural layers. The Orchestration Layer manages global states, coordinates agents and enforces budgetary constraints. The Specialized Agents Layer executes targeted exploitation strategies tailored to each vulnerability classes. The Foundation Layer provides common services such as hybrid payload generation, persistent memory, browser-based verification, and endpoint discovery/reconnaissance. The overall architecture is illustrated in \ref{fig:architecture}.

\subsection{Orchestration Layer}
The Orchestration Layer manages the progression of a scan from initial reconnaissance through multi-step exploitation. Unlike traditional scanners, which treat each vulnerability class as an isolated test, AWE maintains a global exploitation context capturing the evolving state of the adversary. This includes information such as discovered inputs, observed server transformations, authentication status, prior payload attempts, and successful exploitation steps. By storing this information in a unified state model, the orchestrator can reason about how new findings to adaptively influence the overall strategy. For example, upgrading to authenticated testing once credentials are obtained or suppressing redundant payload attempts based on previously observed failures.

At the center of this layer is the Intelligent Orchestrator , which mediates all interactions between components. It collects reconnaissance results, assesses the viability of different vulnerability classes, and selects appropriate agents to invoke. The selection process is done by a LLM that converts reconnaissance output into a prioritized execution plan. Rather than generating payloads directly, the LLM functions as an advisory mechanism, interpreting contextual cues such as reflected parameters, sanitization behavior or the presence of language specific templating constructs. The orchestrator therefore avoids the inefficiency of enumerating every agent and instead executes a minimal subset who meet the required preconditions.

The Orchestration Layer finally also enforces resource governance by monitoring token spend, runtime, and tool costs. This information steers scheduling so AWE can exit early after high-impact findings or scale back low-yield agents, keeping operations within practical limits while focusing depth where it matters.

\subsection{Specialized Agents Layer}
The Specialized Agents Layer embodies the domain knowledge required to navigate specific vulnerability classes. Each agent is implemented as a self-contained exploitation module that translates application behavior into vulnerability specific hypotheses and tests those hypotheses using structured procedures. Rather than relying solely on LLM reasoning, agents encode expert methodologies directly into their operational pipelines, ensuring predictable and reproducible behavior.

\begin{figure}[t]
      \centering
      \includegraphics[height=0.5\textheight]{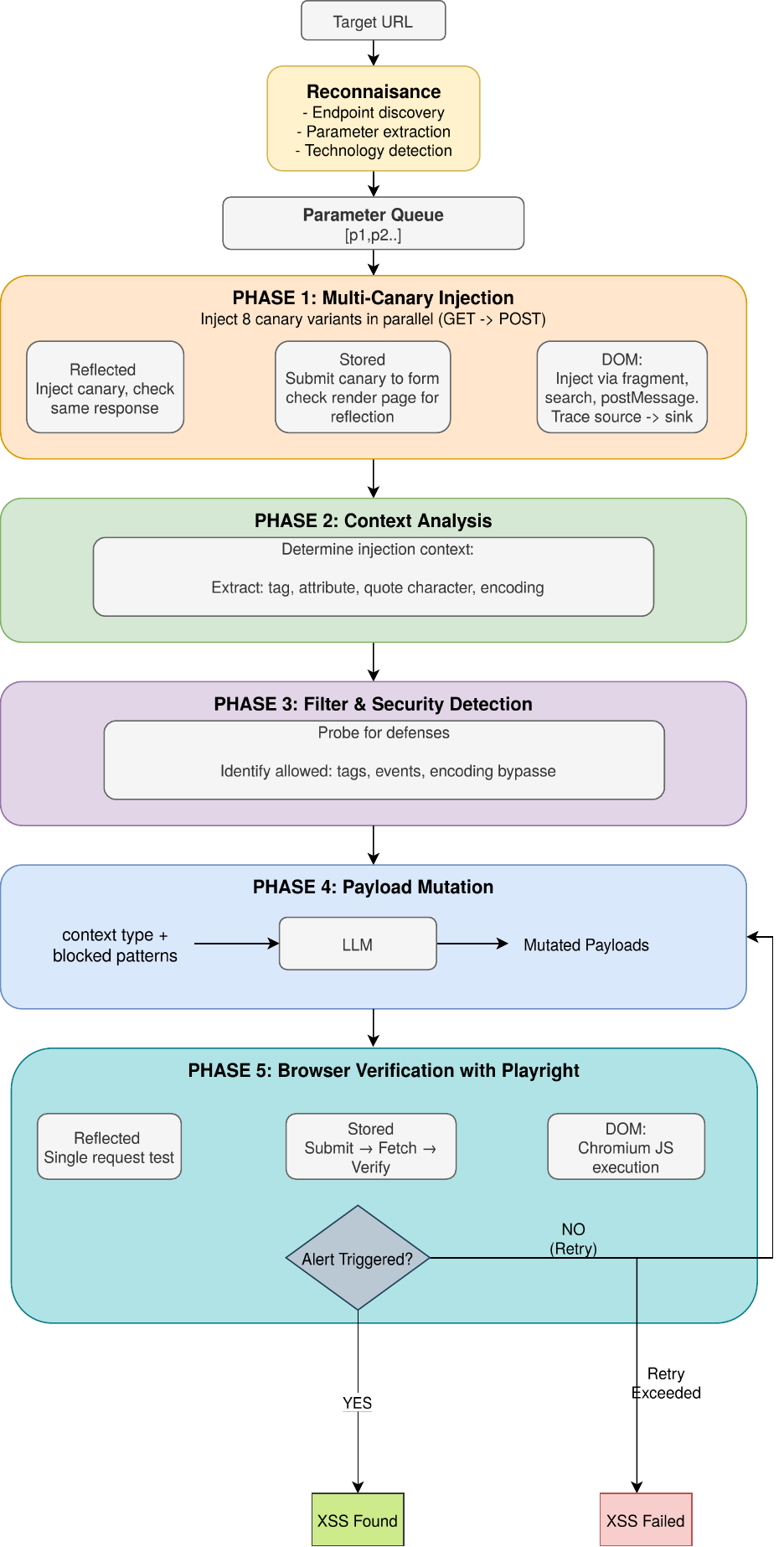}
      \caption{Five-phase XSS detection pipeline.}
      \label{fig:xss_pipeline}
  \end{figure}

The XSS agent is the best example to explain the approach. The agent first conducts a multi-stage analysis beginning with parallel canary injections to map input reflection behavior to analyse for reflected XSS. The agent distinguishes among fine-grained DOM contexts, such as quoted and unquoted attributes, JavaScript string literals, or raw HTML insertion because the viability of subsequent payloads depends critically on contextual correctness. Following context identification, the agent performs targeted probing to infer server side filtering policies, including character level transformations, blocked tag families, and event handler restrictions. This information is packaged into a structured format passed to the LLM, which synthesizes payload candidates tailored to the discovered constraints. This design avoids the speculative or hallucinated vulnerabilities common in LLM-driven systems by grounding LLM creativity in precise contextual information and requiring definitive evidence of JavaScript execution. A visualization of this workflow is shown in \ref{fig:xss_pipeline}.

Agents for SQL injection, server-side template injection, command injection, XXE, SSRF, IDOR, and LFI follow similar principles. The SQL injection agent combines deterministic payload sets with context inference derived from database error messages, backend fingerprinting, and observed query-structure patterns. It then applies controlled mutations to explore alternative execution paths or WAF bypasses. The SSTI agent deploys engine-specific probes to distinguish among popular templating frameworks and subsequently constructs exploit strings that reflect the internal semantics of the detected engine. The IDOR agent relies on authenticated differential testing, comparing access patterns across resource identifiers to detect authorization inconsistencies. In every case, the agents share the same design philosophy: integrate structured domain knowledge, reduce reliance on unconstrained LLM reasoning, and validate exploitability through concrete behavioral evidence.

\subsection{Foundation Layer}
The Foundation Layer provides a shared infrastructure which all agents use and operate upon. 
A core component is the Persistent Memory System, which combines short term scan states with long-term cross-target learning. Short-term memory prevents redundant attempts within the same engagement by tracking tried payloads and their outcomes, inferred filters, and agent-level progress markers. Long-term memory records domain-level features such as effective bypass patterns, characteristic sanitization signatures, and historical payload success rates. This architecture allows AWE to integrate prior experience into future attacks and reduces unnecessary exploration, mirroring how a real adversary accumulates knowledge over repeated interactions with similar systems
The Browser Verification Engineis another core component provides definitive exploit validation for vulnerability classes whose manifestation cannot be confirmed through HTTP responses alone. By executing payloads in a controlled browser environment, AWE observes concrete signals script execution, DOM mutation, dialog triggers that are otherwise invisible to purely server-side testing. This eliminates entire categories of false positives and differentiates between theoretical and practically exploitable vulnerabilities.

Finally, services such as endpoint discovery, parameter extraction, and technology fingerprinting populate the initial attack surface and inform orchestrator decision-making. These components ensure comprehensive yet efficient enumeration of reachable interfaces and reduce unnecessary agent invocation by identifying the structural features most relevant to specific vulnerability classes.

\subsection*{Design Rationale}
AWE’s architecture reflects three principled design choices:

\begin{enumerate}
    \item \textbf{Specialization over generalized reasoning.}
    While LLMs excel at reasoning about semi-structured tasks, fine-grained exploitation requires
    domain-specific procedures that are more reliably implemented as dedicated state machines
    and inference pipelines.

    \item \textbf{Stateful and memory-driven operation.}
    Modern exploitation depends on multi-step reasoning that spans numerous requests, input
    transformations, and contextual clues; a stateless scanner or unconstrained LLM agent cannot
    reliably maintain such context.

    \item \textbf{Verification rather than speculation.}
    Every finding in AWE must be supported by concrete evidence---observable execution, differential
    behavior, or successful data extraction---ensuring that the system reports only vulnerabilities
    that a real adversary could exploit.
\end{enumerate}

These design principles collectively enable AWE to operate as a practical, resource-bounded,
and exploit-grounded autonomous penetration tester capable of discovering complex web
vulnerabilities with high precision.

\section{Methodology}
This section outlines our evaluation methodology, including benchmark selection, baselines, model experiments, configuration, and metrics. The goal is to assess AWE’s effectiveness and efficiency under realistic attacker constraints while enabling reproducible comparison with state-of-the-art systems.
\begin{figure}[t]
\centering
\includegraphics[width=0.9\columnwidth]{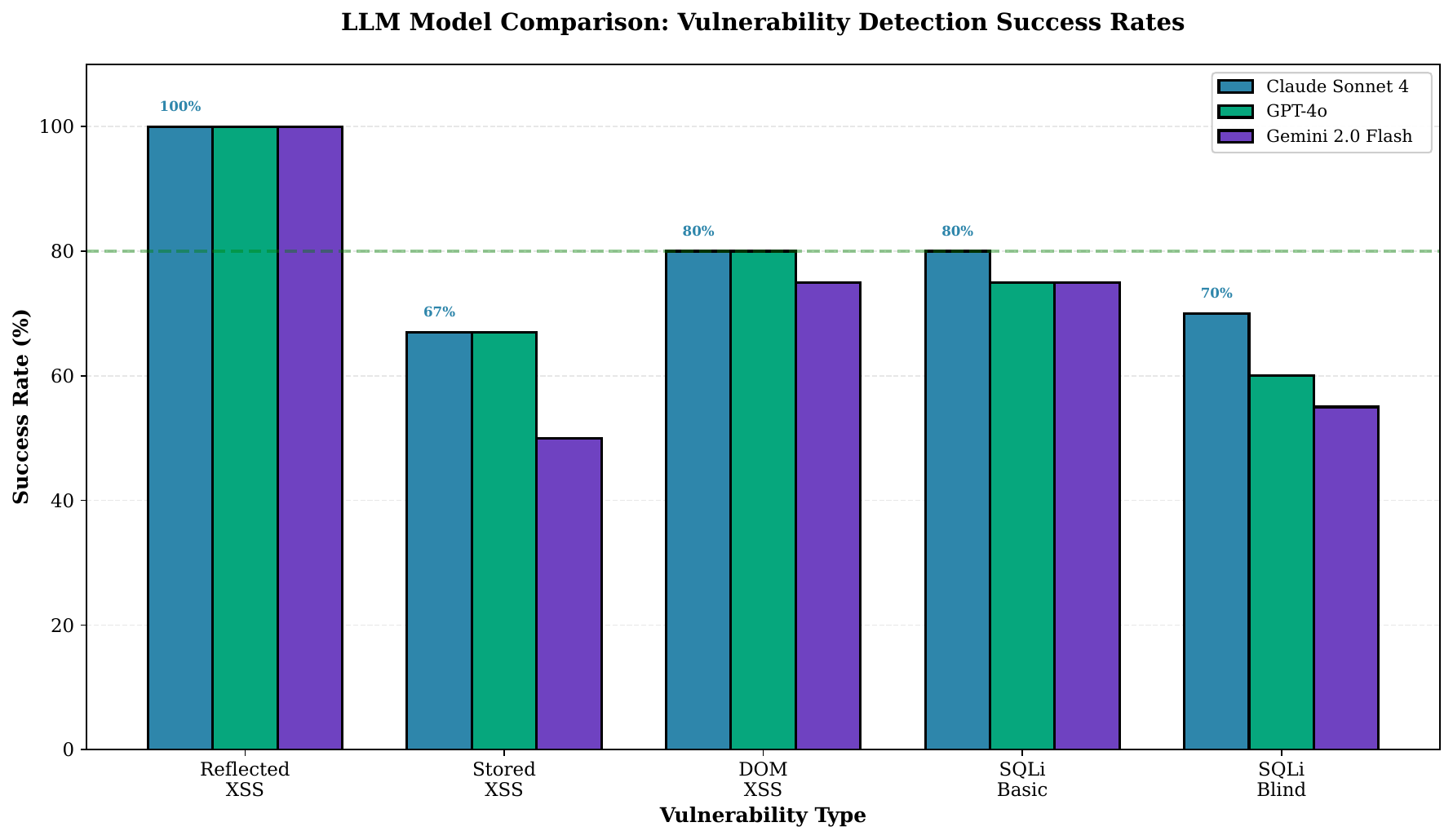} 
\caption{Comparative performance of Claude Sonnet 4, GPT-4o, and Gemini 2.0 Flash across five vulnerability categories.}
\label{fig2}
\end{figure}

\subsection{Benchmarks}
We evaluate AWE on two complementary benchmarks to assess both competitive performance and controlled vulnerability analysis.
\subsubsection*{XBOW Benchmark}
Our primary evaluation uses the XBOW benchmark \cite{xbow}, a curated suite of 104 vulnerable web applications spanning 26 vulnerability categories. Each challenge is deployed as an isolated container and embeds a hidden flag that is accessible only through a complete end-to-end exploit. XBOW provides substantial heterogeneity: vulnerabilities range from straightforward reflected XSS to multi-stage chains involving authentication, authorization, and context-specific sanitization bypasses. Injection-related categories constitute a majority of the benchmark mimicking the state of real world vulnerabilities.
Challenges differ in exploitation complexity. Some are solvable through single-step injections, whereas others require the adversary to combine multiple findings, sequence authenticated requests, or adapt payloads to nontrivial server-side filters. This diversity makes XBOW a suitable testbed for evaluating AWE’s ability to perform adaptive exploitation at scale.
\subsubsection*{DVWA}
For controlled model-selection experiments and fine-grained analysis of exploitation behavior, we use DVWA (Damn Vulnerable Web Application) \cite{dvwa}. DVWA offers repeatable vulnerability configurations and configurable security levels, enabling systematic testing across multiple difficulty regimes. We focus on reflected and stored XSS, DOM-based XSS, error-based SQL injection, and time-based blind SQL injection. Because the application is deterministic across runs, DVWA supports statistical comparison of model behavior under identical conditions. Each model is evaluated across multiple independent trials (n=10) per vulnerability type to obtain robust estimates of success rates and convergence behavior. 
\subsection{Baseline}
We compare AWE against MAPTA in the XBOW Benchmark as it is the strongest publicly available autonomous penetration-testing framework. MAPTA adopts a general purpose multi-agent architecture in which a central LLM orchestrates reconnaissance, execution within an isolated sandbox, and exploit validation. MAPTA’s published evaluation reports a 76.9\% solve rate on XBOW under generous compute and time budgets. Its architecture embodies the prevailing paradigm of broad, reasoning-centric agents, making it an appropriate baseline for measuring the benefits of AWE’s specialization-oriented design. We use MAPTA’s publicly reported per-challenge results for all comparisons.

\subsection{Model Selection}
Before large-scale evaluation, we compared several modern LLMs within AWE’s orchestration layer using DVWA. Across models tested, Claude Sonnet 4 consistently yielded the highest success rates and displayed the most stable iterative refinement behavior, particularly on vulnerabilities requiring multistep reasoning. The detailed numerical results appear in \ref{subsec:dvwasub}, where we revisit this analysis alongside full evaluation. Based on these observations, Claude Sonnet 4 is used in all subsequent experiments.

\subsection{Experimental Configuration}
AWE is evaluated in its aggressive configuration, which performs deep reconnaissance and executes all agents deemed relevant by the orchestrator. Each challenge is allotted a ten-minute time budget, matching MAPTA’s configuration to ensure comparability. All experiments execute on identical hardware, and each challenge is run in an isolated environment to prevent cross-contamination. Memory state is reset between challenges to evaluate single-engagement performance, and browser-based verification is performed using a consistent, headless Chromium configuration. 

\subsection{Evaluation Metrics}
We assess AWE along three principal dimensions: effectiveness, efficiency, and cost. Effectiveness is measured using overall and per-category solve rates, as well as the number of challenges uniquely solved by AWE or MAPTA. Efficiency metrics include time-to-solve and token usage per successful exploit, capturing both responsiveness and resource requirements. Cost metrics reflect total API expenditure and amortized cost per solved challenge based on provider pricing.

\subsection{Sucess Criteria}
A challenge is considered solved only if AWE retrieves the correct flag through a verified exploit. Partial progress or vulnerability detection without successful exploitation is not counted toward effectiveness metrics. This strict criterion ensures that all reported successes correspond to practically realizable attacks.

  \begin{table}[t]
  \centering
  \caption{Overall performance on the XBOW benchmark.}
  \label{tab:overview}
  \begin{tabular}{lccc}
  \toprule
  \textbf{System} & \textbf{Solve Rate} & \textbf{Avg. Time (s)} & \textbf{Model} \\
  \midrule
  AWE   & 51.9\% (54/104) & 53.1  & Claude Sonnet 4 \\
  MAPTA & 76.9\% (80/104) & 190.8 & GPT-5 \\
  \bottomrule
  \end{tabular}
  \end{table}

\section{Evaluation}
We evaluate AWE through a two-stage methodology. First, we conduct controlled experiments on DVWA to isolate the contribution of the underlying language model and justify our choice of Claude Sonnet 4. Second, we benchmark AWE against MAPTA, the most diverse publicly documented autonomous penetration-testing system, on the full 104-challenge XBOW benchmark.

This combination of controlled and large-scale testing provides a comprehensive view of AWE’s capabilities, limitations, and efficiency.

\subsection{Evaluation on DVWA}
DVWA provides a stable, deterministic environment that enables fine-grained comparison of LLM behavior independent of broader architectural factors. We executed AWE with three LLMs - Claude Sonnet 4, GPT-4o, and Gemini 2.0 Flash using identical agent logic and verification procedures across five representative vulnerability classes.
Across all models, reflected XSS served as a baseline of capability, with each model achieving 100\% success. Performance diverged sharply, however, once contextual reasoning or iterative inference became essential. Claude Sonnet 4 consistently outperformed both GPT-4o and Gemini on stored XSS with CSP enforcement and blind SQL injection the two categories that require AWE’s most complex reasoning loops. For CSP-enforced stored XSS, Claude and GPT-4o tied at 67\% accuracy, whereas Gemini dropped to 50\%. For blind SQLi, Claude reached 70\%, GPT-4o 60\%, and Gemini 55\%. These gaps reflect model-dependent differences in temporal inference, semantic constraint handling, and multi-step payload refinement.
We also examined iteration efficiency. Claude converged in 10–40 payload attempts, whereas GPT-4o required roughly 20\% more attempts and Gemini nearly 40\% more. Given that AWE performs many such cycles for complex vulnerability classes, convergence stability directly affects time and cost. Collectively, these DVWA results demonstrate that Claude Sonnet 4 provides the best balance of accuracy and reasoning efficiency. For this reason, all subsequent experiments use Claude Sonnet 4 as AWE’s underlying model. 
\label{subsec:dvwasub}

\subsection{Evaluation on XBOW Benchmark}
We now evaluate AWE on the XBOW benchmark, a suite of 104 containerized web challenges spanning 26 vulnerability categories, ranging from single-step injections to multi-stage exploitation workflows. We use MAPTA as a baseline because it represents the most capable peer system: it employs GPT-5 in extended-reasoning mode and executes arbitrary code within a sandbox, enabling broad exploration beyond what AWE’s specialized agents support.

\subsubsection*{Overall Performance}
Table \ref{tab:overview} presents aggregate results. MAPTA attains a higher solve rate (76.9\%) than AWE (51.9\%), reflecting the advantage of its general-purpose, unrestricted sandbox execution. Despite this, AWE exhibits dramatic efficiency advantages. AWE’s average solve time is 53.1 seconds, which is much faster than MAPTA’s 190.8 seconds. AWE’s total token usage is 1.12M compared to MAPTA’s 54.9M, a 98\% reduction. Correspondingly, AWE’s total API cost is \$7.73 versus MAPTA’s \$21.38, despite MAPTA running on a substantially more capable model. These efficiency gains demonstrate that a specialization-oriented architecture can deliver orders-of-magnitude improvements in operational cost and latency.

\begin{figure}[t]
\centering
\includegraphics[width=0.9\columnwidth]{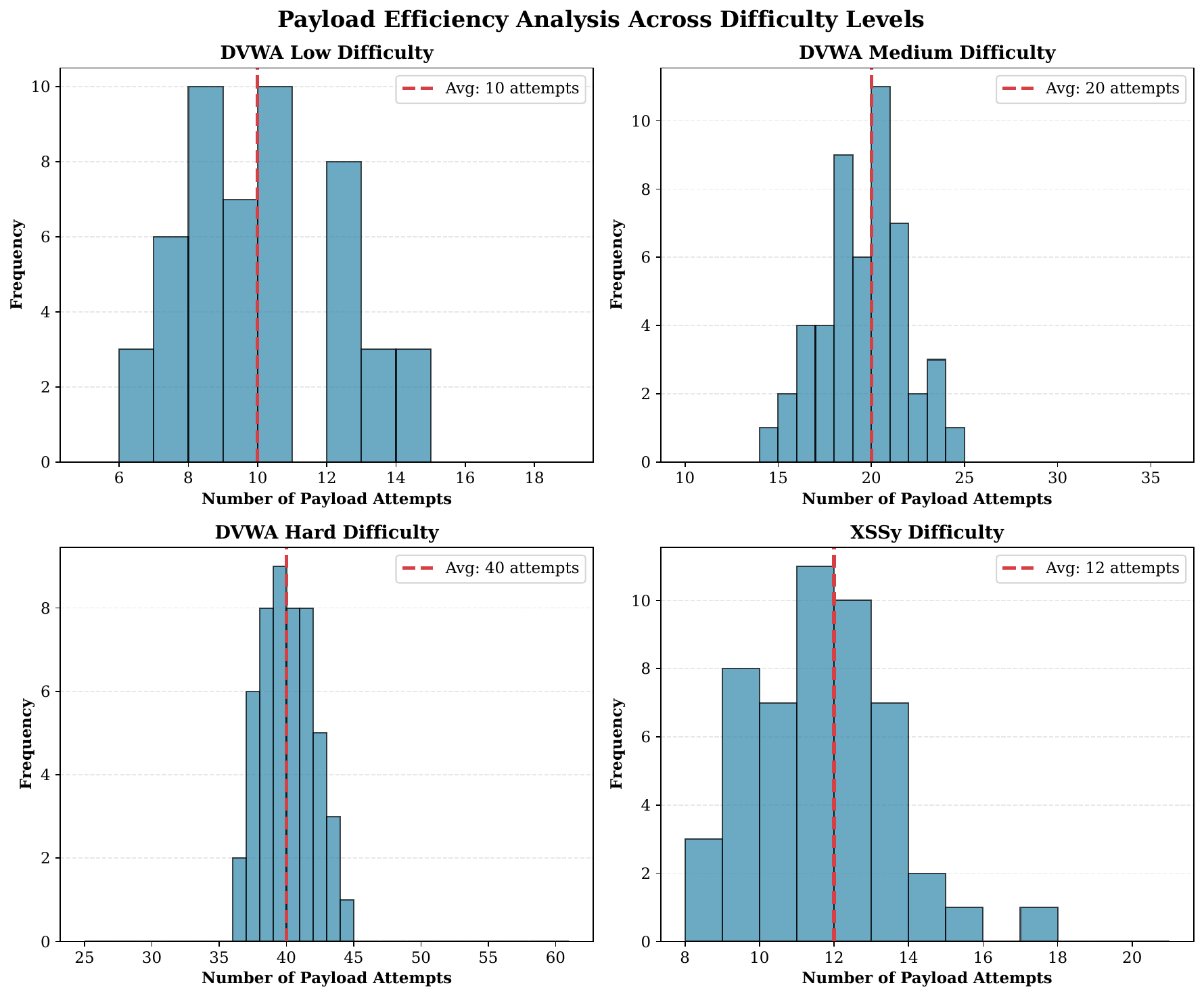} 
\caption{Average number of payload iterations required for successful exploitation by each model. Claude Sonnet 4 converges in the fewest attempts (10–40), followed by GPT-4o with about 20\% more iterations and Gemini 2.0 Flash with about 40\% more. This demonstrates Claude’s superior efficiency and reasoning stability.}
\label{fig3}
\end{figure}

  \begin{table}[t]
  \centering
  \caption{Cost and token efficiency comparison.}
  \label{tab:efficiency}
  \begin{tabular}{lcccc}
  \toprule
  \textbf{System} & \textbf{Total Cost} & \textbf{Cost/Solve} & \textbf{Tokens} & \textbf{Tokens/Solve} \\
  \midrule
  AWE   & \$7.73  & \$0.113 & 1.12M  & 20.7K \\
  MAPTA & \$21.38 & \$0.267 & 54.87M & 685.9K \\

  \bottomrule
  \end{tabular}
  \end{table}

\subsection{Per Category Comparison}
A finer-grained analysis reveals complementary strengths. Table~\ref{tab:vuln} summarizes
category-wise results for the main vulnerability classes. AWE dominates on the injection
classes it explicitly targets, while MAPTA performs better on tasks requiring multi-step
reasoning or semantic exploration. Notably, the two systems perform comparably on the
classical injection families - SQLi, blind SQLi, and XXE - where both models reliably detect
straightforward exploitation patterns.

AWE’s strongest result appears in XSS. Across 23 challenges, it solves 20, substantially
surpassing MAPTA’s 13. The XSS cases solved exclusively by AWE typically require precise
alignment between payload structure and the reflection context (e.g., attribute-versus-
string contexts), adaptive filter bypassing based on observed responses, and reasoning
about multi-encoding transformations. MAPTA’s general-purpose reasoning pipeline often
failed to infer these context-specific constraints.

AWE also performs well on blind SQL injection due to its structured inference workflow
and backend-specific timing probes. Conversely, MAPTA substantially outperforms AWE in
categories involving long-horizon procedural reasoning, such as privilege escalation,
insecure deserialization, and business logic flaws. These tasks exceed the current
capabilities of AWE’s specialized-agent design. 
\begin{table}[t]
\centering
\caption{Category-wise performance comparison on XBOW for Injection Vulnerablities.}
\label{tab:vuln}
\begin{tabular}{lrrrrrr}
\toprule
\textbf{Vulnerability} & \textbf{Total} &
\multicolumn{2}{c}{\textbf{MAPTA}} &
\multicolumn{2}{c}{\textbf{AWE}} \\
\cmidrule(lr){3-4} \cmidrule(lr){5-6}
 &  & \textbf{Count} & \textbf{\%} & \textbf{Count} & \textbf{\%} \\
\midrule
XSS                & 23 & 13 & 57\%  & 20 & 87\%  \\
Blind SQLi         & 3  & 1  & 33\%  & 2  & 67\%  \\
SQLi               & 6  & 6  & 100\% & 6  & 100\% \\
XXE                & 3  & 3  & 100\% & 3  & 100\% \\
SSRF               & 3  & 3  & 100\% & 3  & 100\% \\
SSTI               & 13 & 11 & 85\%  & 7  & 54\%  \\
Command Injection  & 11 & 9  & 82\%  & 5  & 45\%  \\
\bottomrule
\end{tabular}
\end{table}

\subsection{Failure Modes}
AWE failed on 50 challenges; MAPTA failed on 24; both systems failed on 15. Categorizing AWE’s failures reveals that one-third correspond to vulnerability classes intentionally outside its scope (e.g., business logic, deserialization, cryptographic misuse). Another quarter required multi-step reasoning and stateful exploitation chains that AWE’s agents currently cannot express. The remainder fall into authentication irregularities, heavy filtering that resisted AWE’s mutation engine, or extremely narrow exploitation windows (e.g., race conditions).

Challenges solved only by AWE primarily fall into XSS and blind SQLi, reaffirming that its specialized exploitation pipelines provide meaningful advantages even against a more capable underlying model. Conversely, MAPTA-only solves overwhelmingly cluster in categories requiring exploration, multi-agent state management, and semantic reasoning.

\subsection{Effeciency Analysis}

AWE’s primary strength is efficiency. Over the full benchmark, AWE consumed 1.12M tokens compared to MAPTA’s 54.9M - an approximately 98\% reduction. This efficiency stems from two architectural choices: specialized agents avoid the expansive search spaces characteristic of general-purpose reasoning, and memory-guided heuristics significantly reduce redundant attempts.

Time-to-solve exhibits a consistent 4–5× speedup across percentiles. The median solve time for AWE is 35.7 seconds compared to MAPTA’s 156.2 seconds. These gains demonstrate that targeted vulnerability analysis can dramatically reduce overhead without sacrificing performance on its intended classes.

\subsection{Summary}
Our evaluation highlights a clear architectural trade-off. MAPTA achieves broader coverage due to its highly expressive sandbox and frontier-grade model, enabling multi-step exploitation across numerous vulnerability categories. AWE, in contrast, shows that architectural specialization can outperform general-purpose reasoning by large margins on targeted vulnerability classes, even when using a smaller model. The efficiency benefits - 63\% cost reduction, 4.4× faster solves, and 98\% fewer tokens suggest that specialized systems may be preferable for high-frequency testing and integration into continuous assessment pipelines. At the same time, AWE and MAPTA demonstrate complementary strengths, pointing toward hybrid designs that combine structured domain knowledge with general-purpose semantic exploration.

\section{Discussion}
AWE demonstrates that architectural specialization can materially improve the reliability and efficiency of autonomous vulnerability discovery. Its results highlight a broader observation about LLM-driven security testing: general-purpose reasoning alone is insufficient for precise, context-dependent exploitation, while carefully engineered task structure can compensate for smaller model capacity and dramatically reduce computational overhead.

Across XSS and blind SQLi, AWE’s performance stems from explicit modeling of the execution context reflection positions, sanitization behavior, SQL operator boundaries and conditioning payload generation on these abstractions. These constraints reduce the search space an LLM must navigate and yield more stable exploit synthesis than unconstrained reasoning. That AWE outperforms MAPTA on these tasks, despite using a substantially weaker model, suggests that exploit success depends at least as much on architectural priors as on raw model capability.

At the same time, our evaluation shows that specialization does not replace broad autonomous reasoning. MAPTA’s advantages are pronounced on multi-step exploitation involving authentication workflows, privilege escalation, and semantic business logic. These tasks require long-horizon planning and cross-endpoint state tracking capabilities deliberately outside AWE’s design. The contrasting strengths of the two systems indicate that effective autonomous penetration testing will likely require hybrid architectures that combine structured vulnerability analysis with general-purpose exploratory reasoning.

AWE’s efficiency - 98\% fewer tokens, 63\% lower cost, and 4.4× faster solves suggests immediate applicability in continuous or high-frequency testing settings where general-purpose agents remain prohibitively expensive. The ability to embed domain knowledge into agent design also opens the door for adaptive long-term learning: storing filter signatures, past bypasses, and effective payload patterns may enable stable performance across evolving application landscapes.

\section{Limitations}

AWE’s design introduces several boundaries that shape its current applicability:

\begin{itemize}
    \item \textbf{Scope restrictions.} 
    The system targets injection-centric vulnerabilities and does not attempt reasoning-heavy categories such as business logic, complex authentication workflows, or protocol-level issues (e.g., request smuggling or desynchronization).

    \item \textbf{Limited multi-step planning.}
    AWE’s agents operate in largely independent pipelines and do not coordinate multi-stage exploitation sequences. Tasks requiring chained discovery default credentials $\rightarrow$ IDOR $\rightarrow$ privilege escalation fall outside its reach.

    \item \textbf{Reliance on heuristic abstractions.}
    While effective, AWE’s context and filter models encode assumptions about server behavior and sanitization patterns. Highly idiosyncratic frameworks or obfuscated sinks may invalidate these abstractions.

    \item \textbf{LLM sensitivity.}
    Although Claude Sonnet~4 performed best in our analysis, model-dependent reasoning variability remains a systemic constraint; shifts in model behavior or pricing may affect long-term stability.
\end{itemize}

\section{Conclusion}
This work introduces AWE, a specialized multi-agent system that rethinks how LLMs can support autonomous web exploitation. By embedding domain knowledge into the architecture rather than relying solely on free-form reasoning, AWE achieves high accuracy on targeted vulnerability classes and delivers large efficiency gains over a state-of-the-art general-purpose system. The contrast with MAPTA underscores a central insight: precision exploitation benefits from structure, while broad coverage benefits from flexibility.

A natural direction forward is the integration of these paradigms combining specialized agents that capture the semantics of injection vulnerabilities with higher-level agents capable of planning multi-step attacks. Such hybrid approaches may enable autonomous penetration testing systems that are both scalable and semantically capable, bringing fully automated web security analysis closer to practical reality.

\vspace{12pt}

\end{document}